\documentclass[a4,aps,amsmath,floatfix]{revtex4}
\usepackage{graphicx}
\usepackage{color}
\newcommand{\be}{\begin{equation}}
\newcommand{\ee}{\end{equation}}
\newcommand{\ba}{\begin{array}}
\newcommand{\ea}{\end{array}}
\newcommand{\bea}{\begin{eqnarray}}
\newcommand{\eea}{\end{eqnarray}}
\newcommand{\bdm}{\begin{displaymath}}
\newcommand{\edm}{\end{displaymath}}

\begin{document}

\title{Critical Hysteresis in Random Field XY and Heisenberg Models}

\author{Prabodh Shukla}
\email{shukla@nehu.ac.in}
\affiliation{%
Physics Department \\ North Eastern Hill University \\ 
Shillong-793 022, India}%

\author{R S Kharwanlang}
\affiliation{%
Physics Department \\ North Eastern Hill University \\ 
Shillong-793 022, India}%


\begin{abstract}

We study zero-temperature hysteresis in random-field $XY$ and Heisenberg 
models in the zero-frequency limit of a cyclic driving field. We 
consider three distributions of the random field and present exact 
solutions in the mean field limit. The results show a strong effect of 
the form of disorder on critical hysteresis as well as the shape of 
hysteresis loops. A discrepancy with an earlier study based on the 
renormalization group is resolved.

\end{abstract}

\maketitle

\section{Introduction}

Hysteresis is common in systems subjected to a cyclic force 
~\cite{bertotti}. It means that the response to a changing force depends 
on the history of the force. In particular, the response in increasing 
force is different from that in decreasing force. This is caused by the 
delay in responding to the force. Theoretically hysteresis should 
disappear if the force changes sufficiently slowly but this often 
corresponds to unrealistically long time periods.  Several complex and 
disordered systems such as permanent magnets show hysteresis over the 
longest practical time scales. A microscopic model of this phenomenon 
reveals critical points on the hysteresis loop where non-equilibrium 
susceptibility of the system diverges ~\cite{sethna1,sethna2,dhar}. 
These points are characterized by a diverging correlation length, 
scaling of various quantities, and universality of critical exponents 
that is reminiscent of equilibrium critical phenomena. Universal 
behavior covers a wide class of materials, but aspects of universality 
are best examined in the framework of specific models. We focus on the 
zero temperature hysteresis in classical spin models in a quenched 
random field in the limit of zero frequency of the driving field. These 
restrictions are not so drastic as may appear at first sight. In a 
pioneering work, Sethna et al ~\cite{sethna1} used these simplifying 
features to examine magnetic hysteresis in the random field Ising model. 
Their model reproduces several experimentally observed features 
including the shape of hysteresis loops, Barkhausen noise, and return 
point memory ~\cite{sethna1,sethna2,dahmen}. They employ a Gaussian 
distribution of the random field with mean value zero and standard 
deviation $\sigma$ that plays the role of a tuning parameter in the 
model. They find a critical value $\sigma_c$ such that for $\sigma < 
\sigma_c$, each half of the hysteresis loop has a first order jump in 
the magnetization at some applied field $h$. The size of the jump goes 
to zero as $\sigma \rightarrow \sigma_c$ from below. If $h_c$ is the 
critical field at which the jump vanishes, $\{h_c,\sigma_c\}$ is a 
non-equilibrium critical point showing scaling of thermodynamic 
functions and universality of critical exponents in its vicinity. There 
appears to be a fair amount of experimental support for the predictions 
of this model ~\cite{sethna2} . Silveira and Kardar ~\cite{silveira1} 
generalized this model to $n$-component continuous spins. Using the 
renormalization group approach they showed that the generalized model 
also shows critical hysteresis, and the exponents are independent of $n$ 
above 6 dimensions. The significance of 6 dimensions is that it is the 
upper critical dimension of the model with a Gaussian random field 
~\cite{parisi}. In 6 and higher dimensions the action is adequately 
described by a quadratic term, and higher order terms are irrelevant in 
the renormalization-group sense. The quadratic action can be solved 
exactly and predicts a second order transition as $\sigma \rightarrow 
\sigma_c$ from below. The quadratic action is $n$-dependent but the 
critical exponents are not.

Recently we solved ~\cite{shukla1} a slightly different variant of the 
$n$-component spin model exactly in the mean field approximation based 
on infinitely weak but infinitely long range interactions. Our variant 
used randomly oriented but fixed length random fields and spins. We were 
surprised to find strikingly different results from those of Silveira 
and Kardar~\cite{silveira1} for their quadratic action. For $n=2$ ($XY$ 
model), we found a first order transition and no critical point. We also 
found wasp-waisted hysteresis loops in this case. For $n=3$ (Heisenberg 
model) as well, we did not find a critical point of the familiar type 
but one with rather peculiar criticality. The question arises why the 
predictions of two qualitatively similar models are so different. Is it 
because of our variant of the random field or due to the use of soft 
(variable length) $n$-component spins in the field theoretic formulation 
of the renormalization group. We are now in a position to resolve this 
question. We recover the predictions of the quadratic action of Silveira 
and Kardar if we keep the length of vector spins fixed but allow the 
magnitude of the random field to have a Gaussian distribution. We also 
consider a rectangular distribution of the magnitudes of random fields 
to examine if a distribution with a compact support gives a different 
behavior than an unbounded distribution. Rectangular and Gaussian 
distributions are known to give qualitatively different results in the 
case of Ising spins~\cite{dhar,liu}. However, for $XY$ and Heisenberg 
spins we find these two distributions yield similar results.

\section{Equations of motion}

The model studied in ~\cite{shukla1} is characterized by the 
Hamiltonian,

\be H=-J\sum_{i,j} \vec{S_i}.\vec{S_j} -\sum_{i} 
\vec{h_i}.\vec{S_i}-\vec{h}.\sum_{i}\vec{S_i} \ee

Here $\{\vec{S_{i}}\}$ are spins located at sites $\{i=1,2,\ldots N\}$ 
of a d-dimensional lattice, $\{\vec{h}_i\}$ are quenched random fields, 
and $\vec{h}$ is a uniform applied field; $\vec{S_{i}}$, $\vec{h}_i$, 
and $\vec{h}$ are $n$-component vectors with magnitudes $|\vec{S_i}|=1$, 
$|\vec{h_i}|=a_i$, and $|\vec{h}|=h$ respectively. The vectors 
\{$\vec{h_i}\}$ are randomly oriented but their magnitudes \{$a_i$\} may 
have one of several different distributions. We consider three cases: 
(i) $a_i$ is constant independent of the sites ($a_i=a$), (ii) $a_i$ has 
a Gaussian distribution with standard deviation $\sigma$ centered at the 
origin, and (iii) $a_i$ has a uniform distribution in the interval [$0 
\le a_i \le \Delta$], and zero otherwise. The discrete time dynamics of 
fixed-length spins at zero temperature is given by,

\be \vec{S_i}(t+1) = \frac{\vec{f_i}(t)}{|\vec{f_i}(t)|} 
\mbox{\hspace{.5cm}} \vec{f_i}(t)=J 
\sum_{j}\vec{S_j}(t)+\vec{h_i}+\vec{h} \ee

Here $\vec{f_i}(t)$ is the effective field at site-$i$ at time $t$ and 
the factor $|\vec{f_i}(t)|$ in the denominator ensures that the dynamics 
does not alter the length of the spin vectors. We characterize the 
system by $m(t)=\frac{1}{N}\sum_i \vec{h}.\vec{S}_i(t)$, the induced 
magnetization per spin along $\vec{h}$. In the mean field limit a spin 
$S_i$ interacts with every other spin $S_j$ with strength $J=J_0/N$. Let 
the applied field $\vec{h}$ be along the $x$-axis. Then the equations 
for the evolution of magnetization in $XY$ and Heisenberg model in the 
three cases of the random field distribution mentioned above 
are~\cite{note1}:

\subsection{ Randomly oriented fields of fixed magnitude $a_i=a$.}

If the quenched fields $\{\vec{h}_i\}$ are randomly oriented vectors of 
a fixed length $a$, the equations relating $m(t+1)$ to $m(t)$ are,

\be m(t+1) =\left. \frac{1}{2\pi}\int_0^{2\pi} \frac{\{J_0 m(t) + h\} + 
a \cos \alpha_i}{[ a^2+2 \{J_0 m(t)+h\} a \cos \alpha_i +\{ J_0 m(t) 
+h\}^2 ]^{\frac{1}{2}}} d \alpha_i \right. \mbox{ [XY model: $n=2$] } 
\ee

\be m(t+1) =\left. \frac{1}{2}\int_0^{\pi} \frac{\{J_0 m(t) + h\} + a 
\cos \alpha_i}{[ a^2+2 \{J_0 m(t)+h\} a \cos \alpha_i +\{ J_0 m(t) 
+h\}^2 ]^{\frac{1}{2}}} \sin \alpha_i d \alpha_i \right. \mbox{ 
[Heisenberg model: $n=3$]} \ee

Equation (3) is a slight generalization of equation (9) in reference 
~\cite{shukla1} for $a=1$. One may understand it intuitively as follows. 
The zero temperature iterative dynamics aligns $\vec{S}_i(t+1)$ along 
the effective field $\vec{f}_i(t)$ at site-$i$. The effective field 
$\vec{f}_i(t)$ is the sum of the mean field $\{J_0 m(t)+h\}$ along the 
$x$-axis (the direction of the applied field) and a random field of 
magnitude $a$ making an angle $\alpha_i$ with the $x$-axis. Thus the 
components of $\vec{f}_i(t)$ along the $x$ and $y$ axes are 
$f_{ix}=\{J_0m(t)+h\}+ a\cos \alpha_i$ and $f_{iy}=a \sin \alpha_i$. The 
component of the unit vector $\vec{S}_i(t+1)$ along the $x$-axis 
contributes to the magnetization $m(t+1)$. Its contribution is equal to 
the cosine of the angle that $\vec{f}_i(t)$ makes with the $x$-axis and 
therefore equal to $f_{ix}/\sqrt{f_{ix}^2+f_{iy}^2}$. This explains the 
integrand in equation (3). The integral over $\alpha_i$ amounts to 
taking an average over all sites to get the magnetization of the system. 
Equation (4) may be understood similarly. In this case the random fields 
as well as the spins are three component vectors and we need a polar and 
an azimuthal angle to specify their orientation. The polar angle 
$\alpha_i$ of $\vec{h}_i$ is measured from the $x$-axis which is again 
the direction of the applied field and the magnetization of the system. 
The azimuthal angle is integrated out trivially because the component of 
$\vec{S}_i(t+1)$ along the $x$-axis does not depend on it leaving us 
with equation (4).

\subsection{ Randomly oriented fields with a Gaussian distribution of 
$a_i$}

Randomly oriented fields with a Gaussian distribution of $a_i$ means 
that each Cartesian coordinate of the $n$-component field $\vec{h}_i$ 
has a Gaussian distribution with average value zero and standard 
deviation $\sigma$. The integrals in the equation relating $m(t)$ to 
$m(t+1)$ are performed more conveniently in polar coordinates. We 
integrate over $a_i$ with appropriate normalization and weight factors 
for two and three dimensional integrals respectively to get,

\be m(t+1) =\left. \frac{1}{2\pi\sigma^2} \int_0^{\infty} 
e^{-\frac{a_i^2}{2\sigma^2}} a_i da_i \int_0^{2\pi} \frac{\{J_0 m(t) + 
h\} + a_i \cos \alpha_i}{[ a_i^2+2 \{J_0 m(t)+h\} a_i \cos \alpha_i +\{ 
J_0 m(t) +h\}^2 ]^{\frac{1}{2}}} d \alpha_i \right. \mbox{ [XY: $n=2$]} 
\ee

\be m(t+1) =\left. \frac{1}{\sqrt{2\pi}\sigma^3} \int_0^{\infty} 
e^{-\frac{a_i^2}{2\sigma^2}} a_i^2 da_i \int_0^{\pi} \frac{\{J_0 m(t) + 
h\} + a_i \cos \alpha_i}{[ a_i^2+2 \{J_0 m(t)+h\} a_i \cos \alpha_i +\{ 
J_0 m(t) +h\}^2 ]^{\frac{1}{2}}} \sin \alpha_i d \alpha_i \right. \mbox{ 
[Heisenberg: $n=3$]} \ee

\subsection{ Randomly oriented fields with a rectangular distribution of 
$a_i$.}

The equations for this case are obtained on the same lines as in the 
preceding section by integrating over $a_i$ with a uniform distribution 
in the interval $[0\le a_i\le\Delta]$.

\be m(t+1) =\left. \frac{1}{\pi\Delta^2} \int_0^{\Delta} a_i da_i 
\int_0^{2\pi} \frac{\{J_0 m(t) + h\} + a_i \cos \alpha_i}{[ a_i^2+2 
\{J_0 m(t)+h\} a_i \cos \alpha_i +\{ J_0 m(t) +h\}^2 ]^{\frac{1}{2}}} d 
\alpha_i \right. \mbox{ [XY: $n=2$]} \ee

\be m(t+1) =\left. \frac{3}{2\Delta^3} \int_0^{\Delta} a_i^2 da_i 
\int_0^{\pi} \frac{\{J_0 m(t) + h\} + a_i \cos \alpha_i}{[ a_i^2+2 \{J_0 
m(t)+h\} a_i \cos \alpha_i +\{ J_0 m(t) +h\}^2 ]^{\frac{1}{2}}} \sin 
\alpha_i d \alpha_i \right. \mbox{ [Heisenberg: $n=3$]} \ee

\section{Hysteresis in $XY$ model}

Although the integral in equation (3) can be written in terms of 
elliptic functions of the first and second kind, it has an appealing 
geometrical interpretation~\cite{mirollo,shukla1} in its present form. 
The left-hand-side represents the average projection of a spin along the 
$x$-axis. This is proportional to two forces acting along the $x$-axis 
(i) mean field $\{J_0 m(t)+h\}$ and (ii) a random field "$a \cos 
\alpha_i$". These account for the numerator in the integrand. The 
denominator ensures that the resultant of the mean field and the random 
field vectors is a unit vector. The orientation of the random field 
changes from site to site and the integral over $\alpha_i$ represents an 
average over it. In order to examine the structure of a fixed point 
$m^*$ of equation (3), it is convenient to set $J_0=1$ and $h=0$ so that 
$m^*$ is equal to the mean field. Consider two circles of radii unity 
and $a$ respectively ($a \le 1$) with their centers separated by $m^*$ 
on the $x$-axis as shown in figure (1). The spin lies on the larger 
circle and the random field on the smaller one. The spin making an angle 
$\theta$ with the $x$-axis generally cuts the smaller circle at two 
points. Correspondingly two orientations of the random field, one making 
an angle $\alpha$ and the other $\alpha+\beta$ with the $x$-axis produce 
the same magnetization $m=\cos\theta$. Using the properties of triangles 
in figure (1), we get $\beta+2(\alpha-\theta)=\pi$ and $\sin 
(\pi-\alpha+\theta)/m^*=\sin \theta/a$. These two relations can be 
combined to give,

\be m^2=\left. 1- \left(\frac{a}{m^*}\right)^2 \cos^2\frac{\beta}{2} 
\right.\ee

\begin{figure}[p] 
\includegraphics[width=.75\textwidth,angle=0]{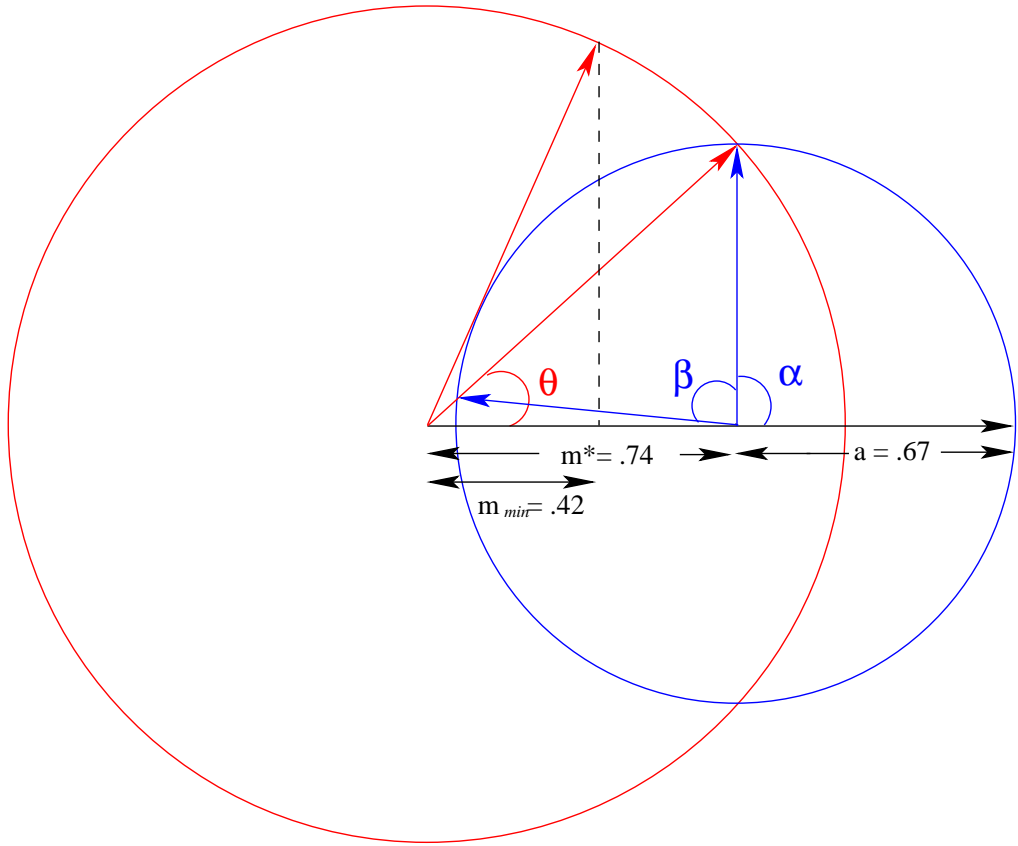}

\caption{ (Color online) The figure shows a geometrical relationship 
between the spin, the random field, and the mean field at the threshold 
of instability in increasing $a$ for $J_0=1$ and $h=0$ (see text).}

\label{FIG1} \end{figure}

In the limiting case $\beta=0$ when the spin vector is tangential to the 
random field circle, $m$ approaches its minimum value $m_{min}$. For a 
given $a$, the fixed point $m^*$ is an average over various values of 
$m$ ranging from $m=m_{min}$ (for $\beta=0$) to $m=1$ corresponding to 
$\alpha=0$ and $\beta=\pi$. Note that the integrand in equation (3) has 
the same value for $\alpha_i=\pi$ as for $\alpha_i=0$. Figure (2) shows 
the fixed points of equation (3) with $J_0=1$ and $h=0$ for increasing 
and decreasing $a$ starting from $m^*=1$ at $a=0$ and changing $a$ in 
small steps $\delta a$. At each step, the fixed point at the preceding 
step is used as a starting point and $a$ is held fixed during the 
iterative dynamics till a new fixed point is reached. We see that $m^*$ 
decreases with increasing $a$; at $a\approx0.67$, $m^*$ jumps down from 
$m^*\approx0.74$ to $m^*=0$ and stays zero thereafter. On decreasing 
$a$, $m^*$ stays zero for $a<0.5$, jumps up to $m^*\approx0.92$ at 
$a=0.5$ and increases along the old trajectory as $a\rightarrow0$. The 
region $0.5 \le a \le 0.671$ is bistable. It contains two lines of 
stable fixed points separated by a line of unstable fixed points. The 
unstable fixed points separate the domains of the stable fixed points. 
In the bistable region, the system is in the domain of $m^*\ne0$ in 
increasing $a$, and $m^*=0$ in decreasing $a$. This is the basic 
mechanism of hysteresis in our model. The boundaries of the bistable 
region can be understood geometrically. As $a$ increases from zero the 
random field circle increases and its center shifts closer to the center 
of the larger circle. The angle $\alpha$ corresponding to $m^*$ also 
increases until $\alpha=\pi/2$ at $m^*\approx0.74$ at $a\approx0.67$.  
This is depicted in figure (1). For larger $a$, $m^*=0$ and the two 
circles are concentric in the stable state. The first order jump in the 
magnetization at \{$h=0,a\approx0.67$\} gives rise to hysteresis loops 
in cyclic fields as shown in figure (3). Note that $m^*=0$ at $h=0$ if 
$a>0.67$ but this does not mean the absence of hysteresis but rather 
wasp-waisted hysteresis loops as explained in reference ~\cite{shukla1}.

\begin{figure}[p] 
\includegraphics[width=.75\textwidth,angle=0]{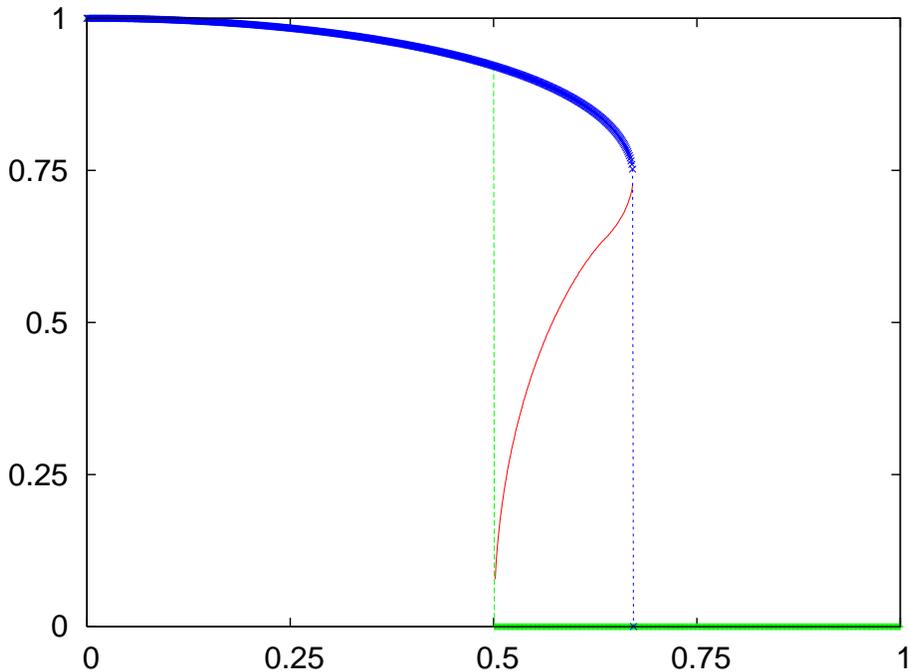}

\caption{ (Color online) Magnetization curves in increasing and 
decreasing $a$ at $h=0$. The thick dark (blue) line shows the 
magnetization in increasing $a$. Starting with $m=1$ at $a=0$, it 
decreases with increasing $a$, drops to zero at $a \approx 0.67$, and 
stays zero for larger values of $a$. For decreasing $a$, it remains zero 
for $a < 0.5$ along the thick gray (green) line, jumps up on the thick 
dark (blue) curve at $a=0.5$ and follows it till the end. The thin dark 
curve (red) shows a line of unstable fixed points in the bistable region 
$0.5\le a \le 0.671$. On decreasing $a$, the starting magnetization is 
always below the line of unstable fixed points, and therefore in the 
domain of the $m^*=0$ fixed point.}

\label{FIG2} \end{figure}

\begin{figure}[p] 
\includegraphics[width=.75\textwidth,angle=0]{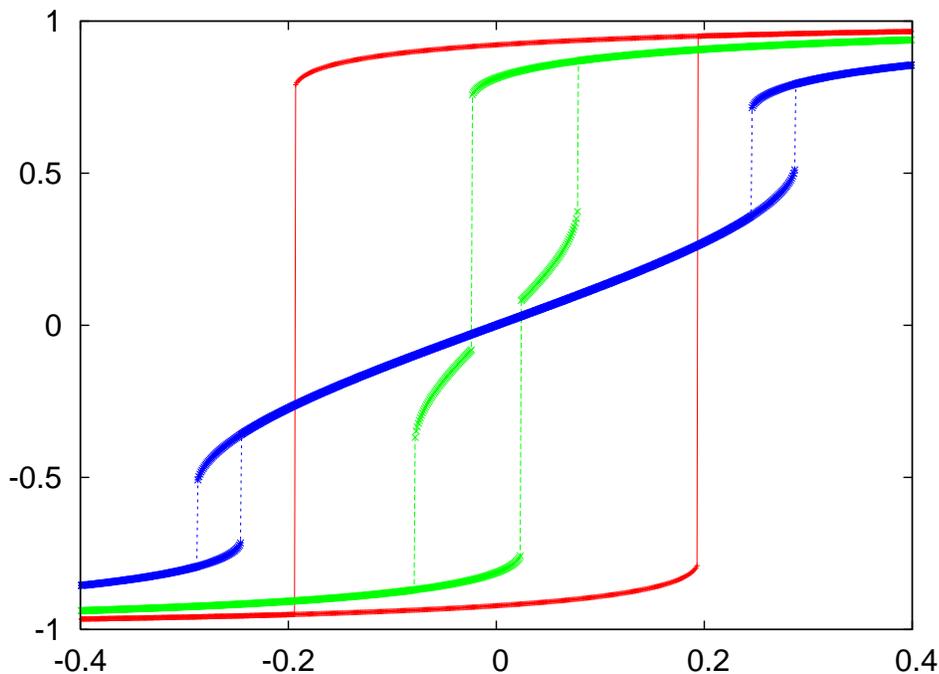}

\caption{ (Color online) Hysteresis loops for the $XY$ model with 
randomly oriented fields of length $a$=0.5 (dark/red line), 0.65 
(thicker gray/green line), and 0.9 (thicker dark/blue line). At 
$a=a_c(\approx0.67)$, $|m(h=0)|$ jumps from $m\approx0.74$ to $m=0$ (see 
figure 2). Therefore hysteresis for $a=0.9$ has a wasp-waisted loop with 
zero hysteresis in small applied fields around $h=0$.}

\label{FIG3} \end{figure}

If the random field at each site has a random orientation $\alpha_i$ and 
a Gaussian distributed magnitude $a_i$ with mean zero and standard 
deviation $\sigma$, our numerical work suggests that the wasp-waisted 
hysteresis loops disappear and the first order jumps in the 
magnetization gradually diminish with increasing $\sigma$ and vanish as 
$\sigma\rightarrow\sigma_c$ from below. We find $\sigma_c\approx0.62$. 
Figure (4) shows the hysteresis loops for two values of $\sigma$, (i) 
$\sigma < \sigma_c$ and (ii) $\sigma >\sigma_c$. The critical behavior 
may be obtained analytically. It is again useful to set $J_0=1$ and 
$h=0$ in equation (5), and then expand it in powers of $m(t)$ in the 
limit $m(t)\rightarrow0$. We split the range of the integral over $a_i$ 
in two parts (i) zero to $m(t)$ and (ii) $m(t)$ to $\infty$ and extract 
the leading terms from both intervals. Thus,

\be m(t+1) = \frac{1}{2\sigma^2}\left[ \int_0^{m_i(t)} 
e^{-\frac{a_i^2}{2\sigma^2}} a_i da_i \left\{2 - \frac{1}{2} 
\left(\frac{a_i}{m_i(t)}\right)^2\right\} +\int_{m_i(t)}^{\infty} 
e^{-\frac{a_i^2}{2\sigma^2}} a_i da_i \left\{ 
\left(\frac{m_i(t)}{a_i}\right) + \frac{1}{8} 
\left(\frac{m_i(t)}{a_i}\right)^3 \right\} \right] \ee

After performing the Gaussian integrals and simplifying, we get

\be 
m(t+1)=\left.\sqrt{\frac{\pi}{8}}\left[\left\{\frac{m(t)}{\sigma}\right\}- 
\frac{1}{8}\left\{\frac{m(t)}{\sigma}\right\}^3 +\ldots \right] 
\right.\ee

This recursion relation shows that $\sigma$ has a critical value 
$\sigma_c=\sqrt{\pi/8}\approx0.6267$; $m^*=0$ if $\sigma > \sigma_c$; 
and $m^*=\sqrt{8\sigma_c} (\sigma_c-\sigma)^{1/2}$ as $\sigma 
\rightarrow \sigma_c$ from below. Thus there is a second order phase 
transition at $\sigma=\sigma_c$. These predictions are born out by 
numerical solution of the equation as shown in figure (5).

For randomly oriented fixed-length fields $\{a_i=a\}$, the corresponding 
equation for the magnetization is,

\be m(t+1)=\left.\frac{1}{2}\left[\left\{\frac{m(t)}{a}\right\}+ 
\frac{1}{8}\left\{\frac{m(t)}{a}\right\}^3 +\ldots \right] \right.\ee

As explained in reference ~\cite{shukla1}, the positive sign of the 
cubic term in the above equation rules out a continuous transition from 
$m^*=0$. There is a first order transition in this case. One may ask if 
the difference in hysteresis for a Delta function distribution 
$\{a_i=a\}$ and a Gaussian distribution comes from the difference 
between a sharply localized distribution and an unbounded distribution. 
In order to examine this question we first relax the equality $a_i=a$ 
and allow $a_i$ to be uniformly distributed in a narrow range 
[$a-\epsilon$ to $a +\epsilon$]. We find that the hysteresis for a 
narrow rectangular distribution around $a_i=a$ is qualitatively the same 
as that for $a_i=a$ in the limit $\epsilon \rightarrow0$. In the same 
vein, we also consider the distribution $a_i=\frac{1}{\Delta}$ if $0 \le 
a_i \le \Delta$, and $a_i=0$ otherwise. For this distribution, equation 
(7) takes the following form in the limit $m(t) \rightarrow 0$,

\be m(t+1)= \frac{m(t)}{\Delta} 
\left[1-\frac{1}{8}\left\{\frac{m(t)}{\Delta}\right\}^2 + \ldots \right] 
\ee

The above recursion relation has a critical point at $\Delta_c=1$; 
$m^*=0$ if $\Delta \ge \Delta_c$, and $m^*=\sqrt{8\Delta^2 
(\Delta_c-\Delta)}$ as $\Delta$ approaches $\Delta_c$ from below. Figure 
(6) shows $m^*(h=0)$ as $\Delta$ increases from zero to $\Delta > 
\Delta_c$. Thus the critical behavior of the model with a uniform 
bounded distribution of $\{a_i\}$ is the same as for a Gaussian 
distribution.

\section{Hysteresis in Heisenberg model}

We now examine hysteresis in the Heisenberg model with each of the three 
distributions considered above. For random field vectors of a fixed 
length $|\vec{a_i}|=a$ and $A(t)=J_0m(t)+h$, equation (4) simplifies to 
the following:

\bea m(t+1)= \frac{2 A(t)}{3 a} \mbox{ if } |A(t)| \le a \nonumber \\ = 
1 - \frac{1}{3} \left\{\frac{a}{A(t)}\right\}^2 \mbox{ if
} A(t) > a
\nonumber \\ =-1 +\frac{1}{3} \left\{\frac{a}{A(t)}\right\}^2 \mbox{ if
} A(t)
< -a \eea

Setting $J_0=1$ and $h=0$, the fixed point in the small $m$ limit is 
determined by the recursion relation $m(t+1)=2m(t)/3a$. Thus $m^*=0$ is 
stable fixed point if $a>2/3$. At $a=2/3$ any value in the range 
$-\frac{2}{3}\le m^* \le \frac{2}{3}$ satisfies the fixed point equation 
at $h=0$. This is a peculiarity of mean field hysteresis in the 
Heisenberg model with randomly oriented fields of fixed length $a$. If 
$a>\frac{2}{3}$, $|m^*|>\frac{2}{3}$ and approaches unity as 
$a\rightarrow0$. Therefore as $h$ is cycled between $-\infty$ and 
$\infty$, we get hysteresis if $a<2/3$, no hysteresis if $a>2/3$, but 
hysteresis does not vanish at $a=2/3$ in any familiar fashion of first 
or second order transition.

If the magnitude of the random field has a Gaussian distribution, the 
appropriate recursion relation in the small $m$ limit is obtained from 
equation (6). Setting $J_0=1$ and $h=0$, we get

\be m(t+1) = \sqrt{\frac{2}{\pi}}\frac{1}{\sigma^3}\left[ 
\int_0^{m_i(t)} e^{-\frac{a_i^2}{2\sigma^2}} a_i^2 da_i \left\{1 - 
\frac{1}{3} \left(\frac{a_i}{m_i(t)}\right)^2\right\} 
+\int_{m_i(t)}^{\infty} e^{-\frac{a_i^2}{2\sigma^2}} a_i^2 da_i 
\left\{\frac{2m_i(t)}{3a_i}\right\} \right] \ee

This simplifies to,

\be 
m(t+1)=\left.\sqrt{\frac{8}{9\pi}}\left[\left\{\frac{m(t)}{\sigma}\right\}- 
\frac{1}{10}\left\{\frac{m(t)}{\sigma}\right\}^3 +\ldots \right] 
\right.\ee

In contrast to equation (14), the above equation shows a square root 
singularity in critical hysteresis at 
$\sigma_c=\sqrt{8/9\pi}\approx0.5319$; $m^*=0$ if $\sigma > \sigma_c$; 
and $m^*=\sqrt{10\sigma_c} (\sigma_c-\sigma)^{1/2}$ as $\sigma 
\rightarrow \sigma_c$ from below. Therefore magnetization curves for 
$\sigma > \sigma_c$ should be reversible and those for $\sigma < 
\sigma_c$ should show hysteresis loops. Figure (4) shows the results for 
$\sigma=0.5$ and $\sigma=0.7$ respectively. Figure (5) shows the 
critical behavior of $m^*(h=0)$ as $\sigma \rightarrow \sigma_c$. The 
fit with the expression derived above is good over a rather wide region 
$\sigma \le \sigma_c$. For comparison, we have also shown the result for 
the Gaussian random field Ising model in figure (5). In the mean field 
theory, the magnetization vanishes with a square root singularity as 
$\sigma \rightarrow \sigma_c$ irrespective of $n$ ($n=1,2,3$) but 
$\sigma_c$ decreases with increasing $n$. This may be expected because 
spins with larger number of components have more freedom to disorder. 
The Gaussian random field Ising model has been investigated extensively 
analytically as well as numerically. Its predictions are in reasonable 
agreement with the experimental observations. It is 
argued~\cite{silveira1} that even for $XY$ and Heisenberg models, if the 
critical hysteresis occurs at $m^*\ne 0$ or $h\ne 0$, it may show 
critical exponents of the Ising model because the non-zero magnetization 
or applied field picks a unique direction in the system. It is not clear 
why the Gaussian distribution should be best suited for comparison with 
experiments, but there is comparatively little study of other 
distributions of the random field. A numerical study shows that critical 
exponents of three dimensional random field Ising model with a Gaussian 
distribution are significantly different from those of the same model 
with a bimodal distribution of random fields~\cite{hartmann}.

\begin{figure}[p] 
\includegraphics[width=.75\textwidth,angle=0]{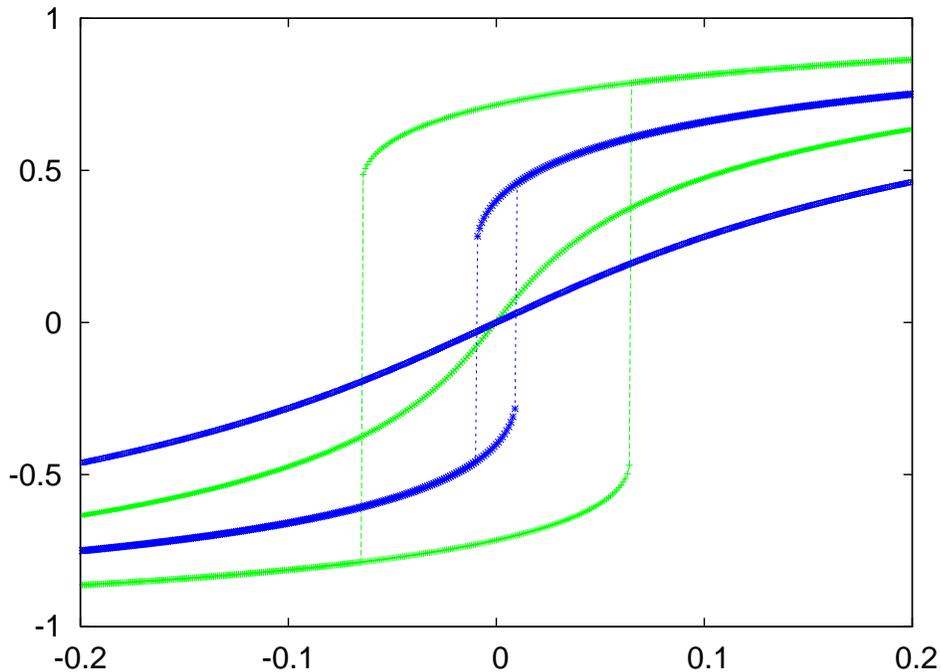}

\caption{ (Color online) Hysteresis in $XY$ and Heisenberg models with 
randomly oriented fields of Gaussian magnitude. Results are shown for 
two values of $\sigma$ in each case; $\sigma=0.5$ ($\sigma < \sigma_c$) 
and $\sigma=0.7$ ($\sigma > \sigma_c$). The gray (green) curves show the 
hysteresis loop and the reversible magnetization for the $XY$ model for 
$\sigma=0.5$ and $\sigma=0.7$ respectively. The dark (blue) curves show 
similar results for the Heisenberg model. The Heisenberg model has a 
smaller hysteresis loop for $\sigma=0.5$ and its reversible 
magnetization for $\sigma=0.7$ has a smaller slope at $h=0$.}

\label{FIG4} \end{figure}

\begin{figure}[p] 
\includegraphics[width=.75\textwidth,angle=0]{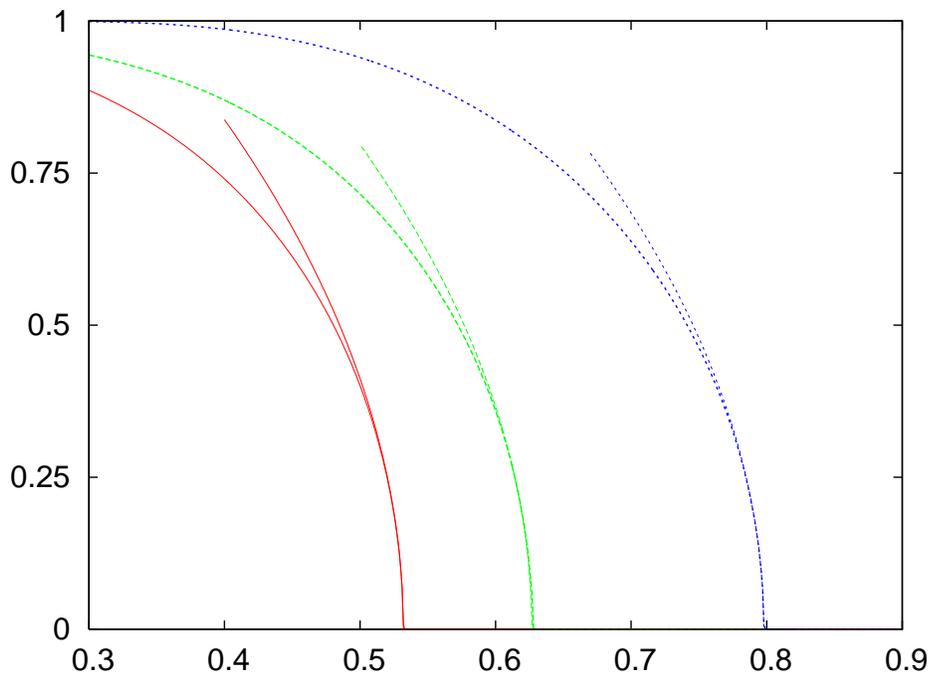}

\caption{ (Color online) Fixed point magnetization $m^*(h=0)$ vs 
$\sigma$ starting with $m=1$: Ising model (the curve to the right/blue; 
$\sigma_c=\sqrt{\pi/2}\approx 0.8$), $XY$ model (the middle curve/green; 
$\sigma_c=\sqrt{\pi/8}\approx 0.6267$), and Heisenberg model (the curve 
to the left/red: $\sigma_c=\sqrt{8/9\pi}\approx 0.5319$). The expression 
$(\sigma_c-\sigma)^{1/2}$ with prefactors $\sqrt{6 \sigma_c}$, 
$\sqrt{8\sigma_c}$, $\sqrt{10\sigma_c}$ (see text) is superimposed on 
each curve respectively. The square root singularity is seen to fit the 
data over a rather wide critical region.}

\label{FIG5} \end{figure}

\begin{figure}[p] 
\includegraphics[width=.75\textwidth,angle=0]{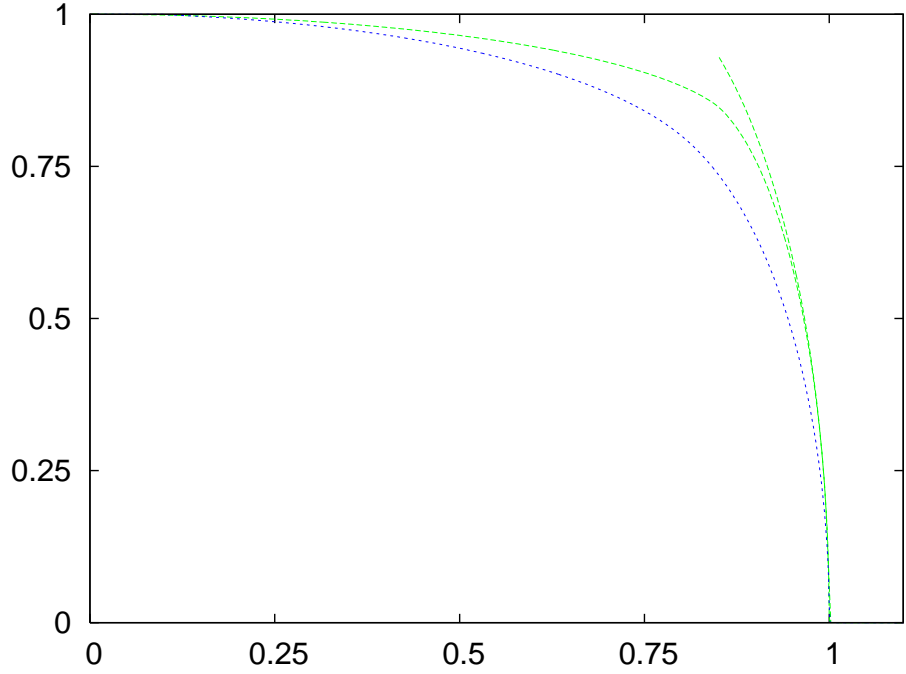}

\caption{ (Color online) Fixed point magnetization $m^*(h=0)$ vs 
$\Delta$ starting with $m=1$: $XY$ model (the gray/green curve; 
$\Delta_c=1$), and Heisenberg model (dark/blue curve: $\Delta_c=1$). The 
expression $\sqrt{8}\Delta(\Delta_c-\Delta)^{1/2}$ is superimposed on 
the $XY$ result to bring out the square root singularity in the critical 
region. For the Heisenberg model $m^*=\sqrt{5\Delta^2(\Delta_c-\Delta)}$ 
fits the curve exactly for $m^* \le \Delta$ (see text).}

\label{FIG6} \end{figure}

In the case of a rectangular distribution of $a_i$, the equations are 
easily integrated. We get,

\bea m(t+1) =\frac{1}{\Delta} m(t) - \frac{1}{5} 
\left\{\frac{m(t)}{\Delta}\right\}^3 \mbox{ if } m(t) \le \Delta 
\nonumber \\ =1 - \frac{1}{5} \left\{\frac{\Delta}{m(t)}\right\}^2 
\mbox{ if } m(t) > \Delta \eea

These equations show that similar to the case of $XY$ model with a 
rectangular distribution of $a_i$, the Heisenberg model shows a critical 
point at $\Delta_c=1$. The fixed point is $m^*=\sqrt{5 (\Delta_c-\Delta) 
\Delta^2}$ if $m^*\le\Delta$ giving a square root singularity similar to 
that for a Gaussian distribution. If $m^*>\Delta$, $m^*$ is determined 
by the real stable root of the cubic equation 
$m^3-m^2+\frac{1}{5}\Delta^2=0$. Figure (6) shows the fixed point 
magnetization at $h=0$ as $\Delta$ is increased from zero to $\Delta > 
\Delta_c$.

\section{Concluding remarks}

This work has been motivated by the need to resolve the discrepancy 
between two exactly solved models of critical hysteresis in 
$n$-component spins ($n=2,3$) placed in a quenched random field. A 
calculation~\cite{silveira1} based on the RG approach predicts a second 
order transition with $n$-independent critical exponents if $d\ge6$. 
This calculation uses a Gaussian distribution of the random field. The 
second calculation~\cite{shukla1} does not involve $d$ explicitly but 
considers the mean field approach of each spin interacting equally with 
every other spin. It employs randomly oriented unit vector fields and 
predicts a first order transition and wasp-waisted loops if $n=2$, and a 
peculiar transition if $n=3$. We have shown that if the random fields in 
the second variant of the model have a Gaussian distribution, the mean 
field theory reproduces the results of the RG approach. This is 
reassuring and may have been anticipated. However an explicit 
verification is valuable because the two models and methods of solution 
are not identical. The RG approach uses variable length spins with 
nearest neighbor interactions on a d-dimensional lattice while the spins 
in the mean field approach are unit vectors. The striking difference in 
hysteresis between Delta function and Gaussian distribution of the 
magnitude of random fields is not generic to a localized and an 
unbounded distribution. We find that a rectangular distribution with a 
compact support yields the same critical behavior as the Gaussian 
distribution in the mean field theory. Thus the calculation presented 
here restores some expectations of the universality of non-equilibrium 
critical behavior but it also raises other questions that we discuss in 
the following. It brings out an unexpectedly strong dependence of the 
shape of hysteresis loops and critical hysteresis on the distribution of 
the quenched field in the system at the mean field level. These results 
have bearings on hysteresis in magnetic systems beyond the mean field 
theory~\cite{sethna1, sethna2, silveira1, parisi,shukla1, dieny, 
silveira2, pierce1, pierce2, jagla} as well as other adiabatically 
driven disordered systems ~\cite{crawford,buscaglia, 
marchetti,saunders,biljakovic,strogatz, gingras,bennett,cardone}. We 
hope the 
analysis presented here may contribute to a better understanding of 
hysteresis experiments and also motivate future work to explore what 
features of quenched disorder are crucial in determining the observed 
phenomena.

It would be interesting to understand our analytic results on general 
grounds such as the symmetry of the system and universality classes of 
critical behavior. Why is it that randomly oriented fixed-length fields 
produce different critical behavior than randomly oriented fields with 
Gaussian distributed magnitudes? Why is it that Gaussian distributed 
magnitudes produce the same critical behavior as uniformly distributed 
magnitudes? We know that critical hysteresis for Ising spins is 
different for a uniform bounded distribution of fields than for a 
Gaussian distribution ~\cite{dhar,liu}. An exact solution ~\cite{dhar} 
of the model on a Bethe lattice with coordination number $z$ reveals 
that there are no jumps in the magnetization nor there is a critical 
point on the hysteresis loop if the fields have a Gaussian distribution 
and $z\le3$. For a uniform bounded distribution of fields there are 
jumps in the magnetization even for $z=3$, and the jumps go to zero 
{\em{discontinuously}} as the width of the distribution is increased. 
This is due to instabilities in the equations of motion at large 
quenched fields on the boundary of the distribution. There is also a 
renormalization group argument ~\cite{liu,dahmen} that for Ising spins 
the crucial feature of the probability distribution of random field is 
its second derivative at zero field. In this formalism the difference in 
the critical behavior of the models with uniform and Gaussian 
distributions is attributed to the difference in the second derivative 
of the probability distribution of the random field at zero field. 
Although the renormalization group analysis as well as an explicit 
solution of equations of motion with a given initial condition lead to 
similar conclusions for critical hysteresis in the case of Ising spins 
but it does not suggest any simple renormalization group argument why 
uniform and Gaussian distributions should yield the same critical 
hysteresis in $XY$ and Heisenberg models. The results presented here are 
based on an exact solution of the equations of motion in the mean field 
approximation. The renormalization group tackles the problem rather 
indirectly.  It recasts the equations of motion into a path integral 
that is a sum over all paths of the exponential of an effective action. 
The effective action carries extra baggage compared with the equations 
of motion. This baggage (extra terms and parameters) comes from the use 
of soft spins and auxiliary fields needed in the exponentiation of delta 
functions. It is expected that the extra terms are irrelevant in the 
renormalization group sense and do not influence the critical behavior 
of the model. However it is not clear to us at this time if the 
difference between a uniform and Gaussian distribution of quenched 
fields is irrelevant for spins and fields having more than one 
component.

An intuitive perspective of results presented in this paper may be 
obtained by paraphrasing them in terms of energy barriers in the 
dynamics of $n$-component spins in quenched fields comprising randomly 
oriented $n$-component ($n\ge2$) vectors of length $a$. Specifically, 
consider the dynamics of the system in the absence of an applied field 
for two different initial conditions (i) all spins pointing along the 
negative $x$-axis ($m_0=-1)$, and (ii) all spins pointing along the 
positive $x$-axis ($m_0=1$). In each case the relaxation dynamics takes 
the system to a stable fixed point whose magnetization is close to the 
initial magnetization if $a<< J$ where $J$ is the ferromagnetic exchange 
coupling between spins. The two fixed points are separated by a large 
gap in their magnetizations and consequently by a large energy barrier 
between them. As $a$ is gradually increased (keeping $J$ fixed) the gap 
between the two fixed points decreases. This is understandable because 
larger $a$ means larger disorder in the system and the fixed point is 
expected to move farther away from the initial uniform state. In the 
mean field theory of $XY$ spins, the gap goes to zero 
{\em{discontinuously}} at a critical value of $a=a_c$. If the lengths of 
the random field vectors $\{a_i\}$ have a uniform distribution in the 
range $0\le a_i\le \Delta$ or a Gaussian distribution in the range $0\le 
a_i \le \infty$ with standard deviation $\sigma$, the gap goes to zero 
{\em{continuously}} at a critical value of $\Delta$ or $\sigma$ as the 
width of the distribution is gradually increased. This too is 
understandable because the distributed magnitudes, particularly very 
small magnitudes in the vicinity of zero magnitude would lower the 
energy barriers for rotation of spins at their sites. The barrier for 
rotation of a spin depends on the magnitude and direction of the 
quenched field at its site as well as the orientation of its neighbors. 
A connected cluster of low $a_i$ sites is likely to have a much lower 
barrier for collective movement in an avalanche than if all $a_i$ were 
equal to each other and of the order of $J$. We may also expect the 
difference between uniform and Gaussian distributions of $\{a_i\}$ to be 
less significant for $XY$ spins than for Ising spins that can only flip 
but not rotate. Similar considerations apply for Heisenberg spins as 
well where the distributed magnitudes $\{a_i\}$ would push the energy 
barriers even lower due to an additional degree of freedom for rotation. 
In a future study we plan to investigate the extent to which the 
predictions of the mean field theory may apply to hysteresis in random 
field $XY$ and Heisenberg models with short range interactions on 
periodic lattices.

\begin{acknowledgments} We thank R Rajesh for useful discussions during 
a visit to NEHU, and Deepak Dhar for a critical reading of the 
manuscript. \end{acknowledgments}


\begin{thebibliography}{99}
\bibitem{note1} These equations are generalizations of corresponding 
equations in ~\cite{shukla1} for $a_i=a$ to different distributions of 
$\{a_i\}$. Also, there is a change of variable. In ~\cite{shukla1} 
$a_i=1$ and $J_0$ is treated as a variable. In the present paper 
$J_0=1$, and $a_i$ is treated as a variable.
\end{thebibliography}
\end{document}